\documentclass[11pt]{article}
\usepackage{comment,soul}
\usepackage[hidelinks]{hyperref}
\usepackage{enumitem}
\usepackage{amsmath,amssymb,epsf,cite,graphicx,subfigure}
\usepackage{amsmath,braket,tikzsymbols}
\usepackage[bbgreekl]{mathbbol}
\usepackage{comment}
\usepackage[export]{adjustbox}
\usepackage{dsfont}
\usepackage{faktor}
\usepackage{mathrsfs}
\usepackage{float}
\usepackage{empheq}
\usepackage{simpler-wick}
\usepackage{mathtools}
\usepackage{array}
\usepackage{multirow}
\usepackage{tabularray}
\usepackage[hmargin=1.5cm]{geometry}
\usepackage[export]{adjustbox}

\numberwithin{equation}{section}

\definecolor{airforceblue}{rgb}{0.36, 0.54, 0.66}
\newcommand{\beq}{\begin{equation}}
\newcommand{\eeq}{\end{equation}}

\newcolumntype{M}[1]{>{\centering\arraybackslash}m{#1}}
\newcolumntype{N}{@{}m{0pt}@{}}

\begin{document}
\baselineskip=15.5pt
\pagestyle{plain}
\setcounter{page}{1}

\begin{center}
{\LARGE \bf A finite Carrollian critical point}
\vskip 1cm

\textbf{Jordan Cotler$^{1,a}$, Prateksh Dhivakar$^{2,b}$, Kristan Jensen$^{2,c}$}

\vspace{0.5cm}

{\it ${}^1$ Department of Physics, Harvard University, Cambridge, MA 02138, USA \\}
{\it ${}^2$ Department of Physics and Astronomy, University of Victoria, Victoria, BC V8W 3P6, Canada\\}

\vspace{0.3cm}

{\tt  ${}^a$jcotler@fas.harvard.edu, ${}^b$pratekshd@uvic.ca, ${}^c$kristanj@uvic.ca\\}

\medskip

\end{center}

\vskip1cm

\begin{center}
{\bf Abstract}
\end{center}
\hspace{.3cm} 
We construct examples of renormalizable Carrollian theories with finite effective central charge and non-trivial dynamics. These include critical points that are not scale-invariant but rather exhibit hyperscaling violation. All of our examples are mildly non-Lagrangian, in that they arise from suitable $N\to 0$ limits of Carrollian theories with $N$-component fields, including limits of Carrollian vector models and non-abelian gauge theories. We discuss implications for flat space holography, highlighting challenges in realizing Carrollian duals to gravitational theories.

\newpage

\tableofcontents

\section{Introduction}

\begin{figure}[h!]
	\centering
	\includegraphics[width=5.7cm]{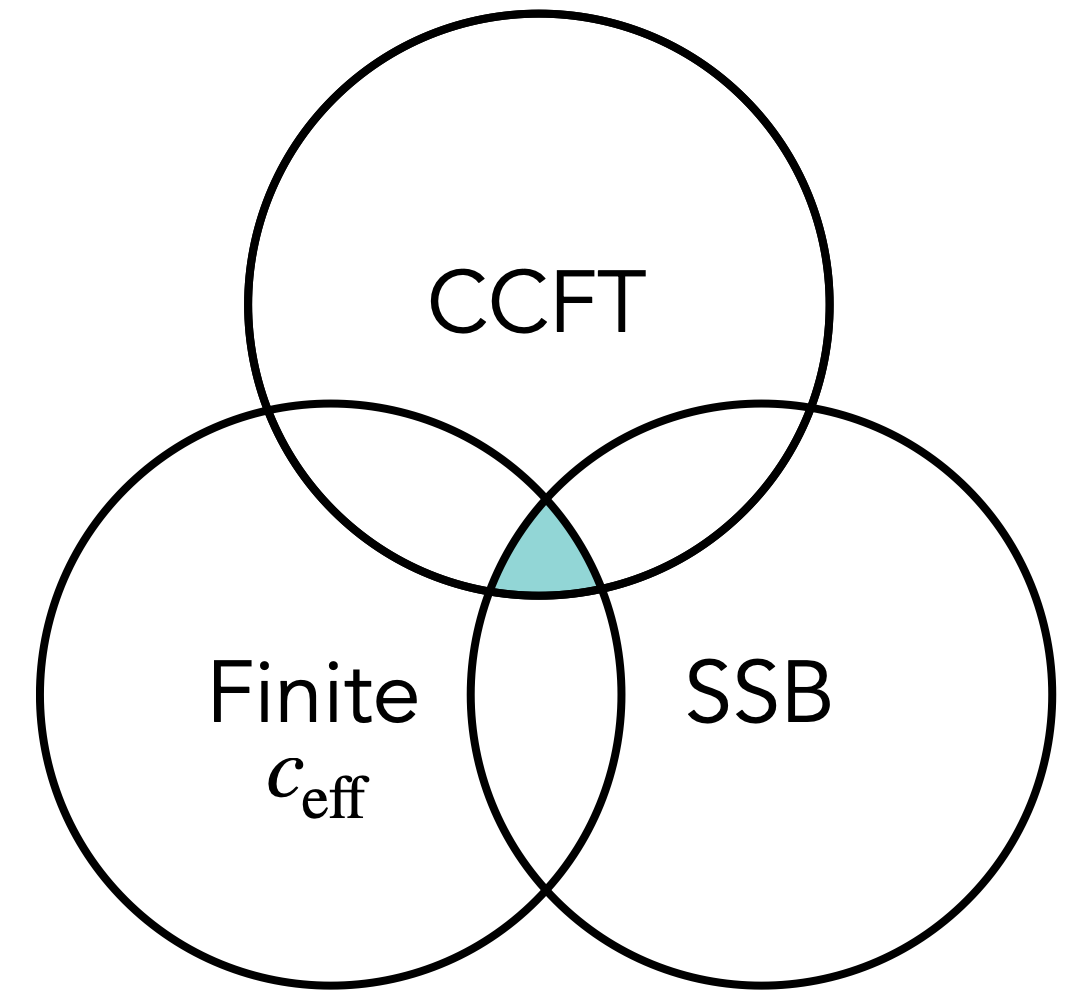}
    \caption{\label{fig:perspective} Features of a Carrollian dual to flat space gravity.}
\end{figure}

Carrollian field theories \cite{Duval:2014uoa,Basu:2018dub,Bagchi:2019xfx,Bagchi:2019clu,Banerjee:2020qjj,Henneaux:2021yzg,Chen:2021xkw,deBoer:2021jej,Hao:2021urq,Bergshoeff:2022eog,Bagchi:2022eav,Saha:2022gjw,Liu:2022mne,Chen:2023pqf,Banerjee:2023jpi,deBoer:2023fnj,Ecker:2024czx,Cotler:2024xhb,Bagchi:2024efs}, which emerge in the ultrarelativistic limit of relativistic field theories, have attracted considerable recent interest as candidate holographic duals to flat space quantum gravity
\cite{Ciambelli:2018ojf,Ciambelli:2018wre,Donnay:2022aba,Bagchi:2022emh,Donnay:2022wvx,Mason:2023mti,Salzer:2023jqv,Nguyen:2023vfz,Bagchi:2023fbj,Saha:2023hsl,Bagchi:2023cen,Alday:2024yyj,Kraus:2024gso,Bagchi:2024gnn,Lipstein:2025jfj}.\footnote{There exists an alternate approach known as Celestial holography. See \cite{Strominger:2017zoo,Pasterski:2021raf,Raclariu:2021zjz} for reviews.} In these proposals, largely inspired by the observation that Poincar\'e symmetry acts on null infinity in the same way as the conformal Carroll group \cite{Duval:2014uva}, $d$-dimensional conformal Carrollian field theories (CCFTs) living on null infinity are envisioned as the dual description of $d+1$-dimensional theories of gravity. Under the current accepted dictionary between the $S$-matrix of a gravitational theory and a putative Carrollian dual \cite{Donnay:2022aba,Bagchi:2022emh,Bagchi:2023cen,Alday:2024yyj,Kraus:2024gso,Jain:2023fxc}, this proposal requires the dual description to satisfy three properties, and likely a fourth as well. The dual must (i) clearly be invariant under the conformal Carrollian symmetries, (ii) have non-trivial $n$-point functions for all $n$, corresponding to bulk scattering processes, and (iii) at least for $d = 2$ and $3$ possess supertranslation and superrotation symmetries which are spontaneously broken to the subgroup of conformal Carrollian symmetries. 
We also expect (iv) that the dual is strongly interacting. See Fig.~\ref{fig:perspective} for a Venn diagram summarizing these constraints.
Each bubble refers to the set of Carrollian theories with that property. The bottom left refers to those that satisfy condition (ii), which in this work we label as ``finite $c_{\rm eff}$'' for reasons that will be clear shortly. The bottom right refers to those theories in which supertranslations/superrotations are spontaneously broken. Viable candidates for a Carrollian dual live in the blue triple intersection.

Recent investigations \cite{Cotler:2024xhb} have revealed several challenges to this possibility, raising the prospect that the blue triple intersection may not exist in the landscape of Carrollian quantum field theories. Two-derivative Carrollian theories have strong sensitivities to both the ultraviolet and the infrared, and commonly exhibit UV/IR mixing.  Of particular interest to the present work, two-derivative Carrollian theories either have no dynamics (so-called ``magnetic theories'') or are models of generalized free fields, which have what we call infinite effective central charge. Under the dictionary mentioned above the former would correspond to a bulk theory with no particles and the latter to one with particles but no interactions. 

The infinite effective central charge implies that the normalization of the stress tensor diverges. To understand this divergence, consider the simplest Carrollian field theory with dynamics, the ultralocal theory of a single scalar field $\phi$,
\begin{equation}
\label{E:prototypeS}
	S = \frac{1}{2}\int du \, d^{d-1}\vec{x} \, \left( (\partial_u \phi)^2 - m^2 \phi^2\right)\,.
\end{equation}
The mixed frequency-position space propagator for $\phi$ is $\langle \phi(\omega,\vec{x}) \phi(-\omega,\vec{0})\rangle = \frac{1}{\omega^2-m^2}\delta^{(d-1)}(\vec{x})$, so that the two-point function of the energy density $\mathcal{E} = \frac{1}{2}\left( (\partial_u \phi)^2 + m^2\phi^2\right)$ which comes from a one-loop bubble is UV-sensitive, going as $\delta^{(d-1)}(\vec{0})$. Similar UV-divergences appear in the correlation functions of composite operators, and when there are cubic or higher-order interactions. 

With others, two of us developed a strategy for treating this UV sensitivity~\cite{Cotler:2024xhb}. In that approach, the first step is to regulate Carrollian field theories by placing them on a spatial lattice with lattice spacing $a$. The second step is to use the lattice spacing to define composite operators and to scale interactions with $a$ in such a way as to admit a continuum theory with a tractable diagrammatic expansion. (One could instead use a spatial momentum cutoff.) As an example in the theory~\eqref{E:prototypeS} above, the operator $\mathcal{O} = a^{\frac{d-1}{2}} \phi^2$ has a finite two-point function in the $a\to 0$ limit. However, $n$-point functions with $n>2$ vanish as $a^{\frac{d-1}{2}(n-2)}$, so these rescaled operators behave like generalized free fields in the continuum limit. The inverse volume of a unit cell $a^{-(d-1)}$ plays the role of a large $N$ parameter controlling the non-Gaussianity of the lattice model, and from it we define an effective central charge $c_{\rm eff} \sim a^{-(d-1)}$.  See also~\cite{Ciambelli:2024swv} for a related, recent discussion.

The stress tensor has a canonical normalization and so the trick above is unavailable. At finite lattice spacing its two-point function comes from a bubble diagram which is given by a frequency integral times a spatial piece which goes as this effective central charge, $a^{-(d-1)}$. The presence of this lattice-sensitive loop factor is an extremely robust feature of Carrollian field theories, not just for simple ultralocal examples like~\eqref{E:prototypeS} but for a family of generalizations studied in~\cite{Cotler:2024xhb}. Indeed a simple tweak of that work demonstrates that all two-derivative, perturbative, and local Carrollian quantum field theories with dynamics have $c_{\rm eff}\sim a^{-(d-1)}$. The argument is very simple. In the quadratic approximation the matrix of propagators of a Carrollian theory has the form \cite{Chen:2021xkw,Bagchi:2022emh,Donnay:2022wvx,Cotler:2024xhb}
\begin{equation}
    \langle \chi_a(x)\chi_b(0)\rangle_0 = E_{ab}(u)\delta^{(d-1)}(\vec{x}) + M_{ab}(\vec{x})\,,
\end{equation}
where $\chi_a$ indexes the fundamental fields and $E_{ab}$, $M_{ab}$ parameterize the propagator. Non-trivial dynamics implies the existence of at least one field with $E_{ab} \neq 0$. Composite operators built from that field which will have UV-sensitive correlation functions going as positive powers of $c_{\rm eff}$.

This result, a weak no-go theorem against the existence of the desired blue triple intersection within the landscape of Carrollian quantum field theories, expresses a Carrollian version of a hierarchy problem, with radiative corrections naturally coming in at the UV cutoff.\footnote{Even supersymmetric theories have the feature that generic operators are naturally UV-sensitive.} More explicitly, there are no finite CCFTs under the basic assumptions of locality, two derivatives, and perturbativity. 

To proceed one must give up certain fundamental properties. Either the dual is a CCFT but of a sort that violates cherished properties like locality; is only accessible as a solution to some suitable version of the bootstrap without any weak coupling limit; or perhaps most radically, the dual is not a field theory in the first place. After all, the proposed flat space holographic dictionary relates the $S$-matrix of a gravitational theory with ``correlation functions'' whose basic requirement is that they transform in the correct way under the action of the conformal Carroll group, not necessarily that they can be interpreted in terms of expectation values of some operators acting on a Hilbert space of a dual description. In this possibility, the dual is more of an engine that produces functions as an output, functions that behave like correlation functions under the action of the Carroll group, but without an underlying quantum mechanical interpretation with Carroll symmetry. 

While the idea of the dual not being a Carrollian field theory may seem initially surprising, precisely the above words describe the flat space limit of AdS/CFT, and with it well-established examples of flat space holography. Recall how the $S$-matrix of a flat space theory arises from a large radius limit of AdS \cite{Penedones:2010ue,Raju:2012zr,vanRees:2022zmr}, equivalently a suitable large $N$ limit of a dual CFT. Given a sequence of holographic CFTs dual to gravity in AdS of arbitrarily large radius, like ABJM with gauge group $U(N)_1\times U(N)_{-1}$ for arbitrarily large $N$~\cite{Chester:2018aca}, one extracts the $S$-matrix of the flat space limit by taking CFT correlation functions, applying a Mellin transform, and then taking a particular large $N$ limit in which one changes the theory, rescales the transformed amplitude by a suitable $N$-dependent constant, and rescales the Mellin variables. In this way the M-theory $S$-matrix is not dual to a single CFT, but to the sequence of ABJM theories as $N\to \infty$. Along the way the CFT data of the sequence is heavily reorganized and the conformal Carroll symmetry acting on null infinity is an emergent property of the limit.

\begin{figure}[t!]
	\centering
	\includegraphics[width=3cm,height=3cm]{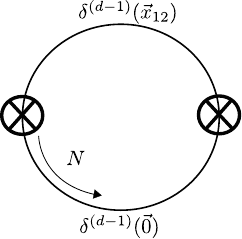}
	\caption{The two-point function of energy density with $N$ fields running around the loop. Identifying $N = \left( \frac{a}{\ell}\right)^{d-1}$ this diagram is parametrically of order $c_{\rm eff} \sim \ell^{-(d-1)}$.}
	\label{fig:oo2point}
\end{figure}

In this manuscript, we obtain finite Carrollian theories by relaxing the assumption of having a field theory, albeit in a particular way in which we seek to deform Carrollian field theories as little as possible. We comment on another possibility below. Our idea is simple. Take the model~\eqref{E:prototypeS}, and consider the singlet sector of its $O(N)$ generalization. Now the two-point function of the stress tensor comes from a vacuum bubble with $N$ fields going around the loop, so that $c_{\rm eff} \sim Na^{-(d-1)}$, the number of degrees of freedom per unit cell of the lattice. See Fig.~\ref{fig:oo2point}. To make $c_{\rm eff}$ finite we analytically continue in $N$, taking $N = \left( \frac{a}{\ell}\right)^{d-1}$ for $\ell$ a length scale, and define a double-scaled version of a continuum limit in which we take $a\to 0$ while holding $\ell$ fixed. The resulting model is a certain $N\to 0$ limit of the Carrollian $O(N)$ model and has a dimensionful effective central charge,
\begin{equation}
    c_{\rm eff} \sim \ell^{-(d-1)}\,.
\end{equation}

The above appears to be the only way to achieve a finite $c_{\rm eff}$ while retaining locality, Carrollian symmetry, and perturbativity. We explore several variants of this idea throughout this work, including an $N\to 0$ version of Carrollian non-abelian gauge theory. This success comes at a cost. The resulting theories are only defined through the diagrammatic expansion, are at best scale covariant rather than scale invariant on account of the length scale $\ell$, and more importantly are not quantum mechanical, since they no longer have a Hilbert space of physical states.

The $N\to 0$ limit of the massless $O(N)$ model has power-law correlations and so describes a critical point. This is the model referred to in the title. This critical point however is not conformal on account of $\ell$. There is really a one-parameter family of theories labeled by the length scale $\ell$, and we can define a spurionic (i.e.~broken) scale transformation which acts as a spacetime dilatation $(u,\vec{x})\to (\alpha u,\alpha \vec{x})$ while changing the theory, $\ell \to \alpha \ell$. In the language of critical phenomena, this critical point violates hyperscaling, with the length scale $\ell$ not decoupling from the infrared. It comes as no surprise that the ensuing critical point is not invariant under Carrollian special conformal transformations either. Under the map from Poincar\'e to conformal Carroll, the bulk boosts become boundary dilatations and special conformal transformations.\footnote{The readers are directed to Appendix B of \cite{Bagchi:2022emh} for a summary of the map.} We conclude that if there was a $d+1$-dimensional dual description then it has broken boost symmetry. 

We study this critical point, its exponents, and generalizations in the main body of this manuscript. In all the cases we study, a finite $c_{\rm eff}$ requires some form of a $N\to 0$ limit like that described above. 

Our approach has been to find Carrollian ``theories'' with finite $c_{\rm eff}$ through certain scaling limits of Carrollian quantum field theories, limits which retain the Carroll symmetry but which sacrifice a quantum mechanical interpretation of correlators. Recently there has been an approach to the problem which shares this feature of giving up on finding Carrollian field theories and instead functioning as an engine for producing Carroll-invariant correlators. Rather than take the $c\to 0$ limit of relativistic CFTs, which leads to field theories with all of the challenges described above and studied in~\cite{Cotler:2024xhb}, the authors of~\cite{Alday:2024yyj} have taken the $c\to 0$ limit of CFT $n$-point functions, allowing themselves the freedom to rescale those correlators by judicious powers of $c$ before taking the limit. Because the relativistic conformal group becomes the conformal Carroll group as $c\to 0$ the resulting functions transform nicely under the conformal Carroll group, and moreover because they insert convenient powers of $c$ before taking the limit, they find finite correlators for all $n$, rather unlike what one finds in the quantization of the Carrollian theory obtained by a $c\to 0$ limit. 

Our interpretation is that the above reflects the perspective of recent works~\cite{Bagchi:2023fbj,Bagchi:2023cen,Alday:2024yyj,Kraus:2024gso} on holographic CFTs, with~\cite{Alday:2024yyj,Kraus:2024gso} further demonstrating that a certain Carrollian limit of Witten diagrams in AdS yields the corresponding Feynman diagrams of the flat space theory emerging in the large radius limit.  See also the very recent work~\cite{Lipstein:2025jfj} that successfully applies this strategy in the context of the AdS$_4\times \mathbb{S}^7$ vacuum of M-theory, thereby extracting tree-level contributions to the M-theory $S$-matrix. 

While these works do not spell out a concrete dictionary between an $S$-matrix and a holographic CFT, ostensibly the idea is that it involves an ultrarelativistic limit of the CFT, thus leading to the desired conformal Carroll symmetry. We note that in order for such a proposal to be viable the large $N$ parameter must be scaled to infinity at the same time, in which case it is not clear to us if such an idea differs from the usual flat space limits of AdS/CFT, beyond giving a dual explanation for the emergence of Poincar\'e symmetry from the $c\to 0$ limit.\footnote{The reader is directed to Section 3.1 of \cite{Bagchi:2023fbj} for a brief history of such flat space limits.}

What do these results imply for the Carrollian version of the flat space holography program? As we mentioned above, a viable dual ought to be conformally invariant, have finite $c_{\rm eff}$, and have the correct supertranslation and superrotation breaking pattern.\footnote{Lagrangian Carrollian theories have a classical supertranslation symmetry that so far appears to be preserved at the quantum level except perhaps in the presence of higher-derivative interactions \cite{Ecker:2024czx,Cotler:2024xhb}, but almost never a classical superrotation symmetry. The only known example of a Carrollian theory with classical supertranslation and superrotation symmetries, both of which are spontaneously broken, is the magnetic theory that arises as the boundary description of pure flat space gravity in three dimensions \cite{Cotler:2024cia}.}  However, we are quite far from finding the desired blue triple intersection of the Venn diagram above, and have established a weak no-go theorem against its existence under certain fundamental assumptions. As far as theories with dynamics go, we have: examples of CCFTs albeit with infinite $c_{\rm eff}$; non-conformal non-field theory examples with finite $c_{\rm eff}$ obtained in this work; no field theories with the correct symmetry breaking pattern; and non-field theory examples obtained from the $c\to 0$ (and presumably also $N\to \infty$) limit of holographic CFTs with finite $c_{\rm eff}$ and only the conformal Carroll symmetry.

We conclude that we must relax fundamental assumptions in order to find viable candidates, whether it be locality, which is perhaps suggested by the proposed gravity/CCFT dictionary,\footnote{According to this dictionary $2\to n$ scattering  with $n\geq 2$ implies certain time-dependent correlations in a dual between spatially separated regions \cite{Banerjee:2019prz,Bagchi:2023cen,Alday:2024yyj}, which suggests the breaking of locality in a Carrollian dual~\cite{Forthcoming}.} some of the postulates of quantum mechanics,  perturbativity, or the restriction to two-derivative theories. If we eschew locality, two derivatives, or perturbativity, the only robust method we have to proceed is a suitable version of the conformal bootstrap.

However, turning to a bootstrap presents new challenges. It is important to clarify which principles should be relaxed and whether, in practice, the resulting bootstrap could realistically produce new solutions. There is also the concern that we might be addressing a problem akin to searching for a needle in a haystack. Moreover, it seems important to note that two holographic duals of the M-theory $S$-matrix, namely the BFSS theory \cite{Banks:1996vh} and a certain $N\to\infty$ limit of ABJM theory~\cite{Chester:2018aca}, are unlikely to be solutions to such a bootstrap given that neither are Carrollian field theories.

The remainder of the paper is organized as follows.  We study the $N\to 0$ limit of the vector model above in Section~\ref{sec:nto0scalar}, and some generalizations in Section~\ref{sec:crits} in which we couple to a magnetic sector, add quartic interactions, or couple to a certain $N\to 0$ limit of Carrollian Yang-Mills. All of these models behave rather similarly to each other.

\section{$N\to 0$ limit of Carrollian vector models}\label{sec:nto0scalar}

\subsection{Setup and basic correlation functions}

In the Introduction, we described an approach to obtain a Carrollian theory with finite effective central charge. Here we study it in some detail.

We begin with the Carrollian version of the free $O(N)$ model,
\begin{equation}\label{eq:freeonact}
	S[\vec{\phi}] = \dfrac{1}{2}\int d^d x \,  \partial_u \vec{\phi} \cdot \partial_u \vec{\phi}  \, ,
\end{equation}
where $\vec{\phi} = \{\phi^I\}, \, I=1,\dots,N$. The two-point function of $\phi^I$ is\footnote{To have an uncluttered notation, we will denote the spacetime dependence $x^{\mu}=(u,\vec{x})$ by $x$.}
\begin{equation}\label{eq:phiphitwopt}
	G^I_{~J}(x_1,x_2) = \langle \phi^I(u_1,\vec{x}_1) \phi_J(u_2,\vec{x}_2)  \rangle = \delta^I_{~J} \, |u_{12}| \, \delta^{(d-1)}(\vec{x}_{12}) \, ,
\end{equation}
where $x_{ij} = x_i -x_j$ and the absolute value ensures that $G$ is a Green's function,
\begin{equation}
	\partial^2_{u_1} G^I_{~J}(x_1,x_2) = \delta^I_{~J} \,\delta^{d}(x_{12}) \, .
\end{equation}
Defining the normal-ordered composite operator
\begin{equation}\label{eq:odef}
	\mathcal{O}(x) = \,: \!\vec{\phi}(x) \cdot \vec{\phi}(x)\! : \, ,
\end{equation}
we find its two-point function
\begin{equation}\label{eq:oo1}
	\begin{split}
		\langle \mathcal{O}(x_1) \mathcal{O}(x_2) \rangle 
		& = \dfrac{2N}{a^{d-1}} u^2_{12} \, \delta^{(d-1)}(\vec{x}_1 - \vec{x}_2) \, ,
	\end{split}
\end{equation}
where we have implemented a lattice regularization on a cubic spatial lattice of lattice spacing $a$. As studied in \cite{Cotler:2024xhb}, the above correlator is divergent in the continuum limit $a\to 0$. Note that taking $N$ to be large will not cure the divergence. Instead, let us consider a more unusual approach, analytically continuing $N = \left( \frac{a}{\ell}\right)^{d-1}$ and holding the length scale $\ell$ fixed while sending $a\to 0$. This defines a dimensionful effective central charge
\begin{equation}\label{eq:nscal2}
	\boxed{c_{\text{eff}} = \ell^{-(d-1)} } 
\end{equation} 
Then using~\eqref{eq:nscal2} in~\eqref{eq:oo1}, we obtain the finite result
\begin{equation}\label{eq:ooini}
	G_{\mathcal{O}}(x_{12},c_{\text{eff}}) = \langle \mathcal{O}(x_1) \mathcal{O}(x_2) \rangle = 2 \,c_{\text{eff}} \, u^2_{12} \,  \delta^{(d-1)}(\vec{x} - \vec{y})  \, .
\end{equation}
This limit entails a continuation of $N$ to non-integer values and taking $N\to 0$, resulting in a ``theory'' in the sense that it produces a set of functions that we can interpret as correlation functions, transform in the expected way under the action of the Carroll group. However such a $N\to 0$ limit breaks the postulates of quantum mechanics, resulting in a model without a Hilbert space of physical states.

An immediate consequence of the scaling of~\eqref{eq:nscal2} is that scale transformations $x^{\mu} \to \alpha x^{\mu}$ are not a symmetry of the theory, even though correlation functions are polynomials in the $u_{ij}$'s and so we have a critical point. Rather there is a spurionic scale transformation under which the effective central charge $c_{\text{eff}}$ also transforms with
\begin{equation}\label{eq:ceffscald}
	c_{\text{eff}} \to \frac{c_{\text{eff}}}{\alpha^{d-1}} \, .
\end{equation}
As we mentioned in the Introduction, this is a signature that this critical point is not scale-invariant but rather exhibits hyperscaling violation. Physically, the length scale $\ell$ does not decouple from the infrared. Using~\eqref{eq:ceffscald} we see that the correlator~\eqref{eq:ooini} is scale covariant under this redefined transformation provided we assign $\mathcal{O}$ the dimension $\Delta_{\mathcal{O}}= d-2$.

We now march on and compute higher-point correlators. For instance, 
\begin{equation}\label{eq:ooo}
	\begin{split}
		\langle \mathcal{O}(x_1) \mathcal{O}(x_2) \mathcal{O}(x_3) \rangle 
		&= 8\, c_{\text{eff}} \, |u_{12} u_{23} u_{31}| \, \delta^{(d-1)}(\vec{x}_{12}) \, \delta^{(d-1)}(\vec{x}_{23}) \, .
	\end{split}
\end{equation}
Inductively, we find that the $n$-point function is
\begin{equation}\label{eq:onpoint}
	\begin{split}
		\left\langle \,\prod_{i=1}^{n} \mathcal{O}(x_i) \right\rangle_{\rm c_{\rm eff}} 
		&= (2n-2)!! \, c_{\text{eff}} \, 
        |u_{12} u_{23} \cdots u_{n-1,n} u_{n1}|\,\delta^{(d-1)}(\vec{x}_{12}) \,\delta^{(d-1)}(\vec{x}_{23}) \cdots \delta^{(d-1)}(\vec{x}_{n-1,n})\,,
	\end{split}
\end{equation}
which is finite in our continuum limit. From this result, we deduce that if we only scale the coordinates without scaling $c_{\text{eff}}$ we get
\begin{align}\label{eq:noceffscalnpoint}
\left\langle \prod_{i=1}^{n} \mathcal{O}(\alpha x_i) \right\rangle_{\rm c_{eff}} = \alpha^{-(d-2)(n-1)-1} \,\left\langle \prod_{i=1}^{n} \mathcal{O}(x_i) \right\rangle_{c_{\rm eff}}\,,
\end{align}
where we have temporarily emphasized the dependence on the effective central charge. Even though the $n$-point function scales homogeneously with $\alpha$, the suite of $n$-point functions do not admit a consistent assignment of a scaling dimension for $\mathcal{O}$. However, instead of \eqref{eq:noceffscalnpoint}, if we scale $c_{\text{eff}}$ as well through \eqref{eq:ceffscald} in addition to dilating space, then we have
\begin{equation}
    \left\langle \prod_{i=1}^{n} \mathcal{O}(\alpha x_i) \right\rangle_{c_{\rm eff}/\alpha^{d-1}} =  \, \left\langle \prod_{i=1}^{n}\left( \alpha^{-(d-2)} \mathcal{O}(x_i) \right)\right\rangle_{c_{\rm eff}}\,.
\end{equation}
So we can consistently assign $\mathcal{O}$ the dimension $d-2$ as anticipated from its two-point function.

Because this critical point is not scale-invariant perhaps it is not a surprise that it fails to be invariant under special conformal transformations either. We have checked this directly for $n$-point functions of $\mathcal{O}$.

The scaling limit we have described does not only lead to finite correlation functions of the composite operator $\mathcal{O}= \vec{\phi}^2$ but of more complicated singlet operators as well without any further redefinition. These include ``multi-trace'' operators built from $\mathcal{O}$, as well as the stress tensor. The (improved) stress tensor has components~\cite{Baiguera:2022lsw}
\begin{equation}\label{eq:stressscalar}
	\begin{split}
		T^u_{~ \, u} &= \dfrac{1}{2} (\partial_u\vec{\phi})^2 \, , ~~~~~~ T^i_{~ \, u} = 0 \, , \\
		T^u_{~ \, i} &= \dfrac{1}{2} \dfrac{d}{d-1} \partial_u \vec{\phi} \cdot \partial_i \vec{\phi} - \dfrac{1}{2} \dfrac{d-2}{d-1} \vec{\phi} \cdot \partial_i \partial_u \vec{\phi} \, , \\
		T^i_{~ \, j} &= -\dfrac{1}{2} \dfrac{d}{d-1} (\partial_u \vec{\phi})^2 \, \delta^i_{~j} + \dfrac{1}{2} \dfrac{d-2}{d-1} \vec{\phi} \cdot \partial^2_u \vec{\phi} \, \delta^i_{~j} \, .
	\end{split}
\end{equation}
and its trace is
\begin{equation}\label{eq:scalartrace}
	T^{\mu}_{~\,\mu} = \dfrac{1}{2}(d-2) \vec{\phi} \cdot \partial^2_u \vec{\phi} \, ,
\end{equation}
which vanishes for on-shell configurations. As an example, the $T^u_{~\,u}$ two-point correlator~\cite{Dutta:2022vkg,Ruzziconi:2024kzo} (i.e. the two-point function of the energy density) is given by a UV-sensitive expression
\begin{equation}\label{eq:tuutwopoint}
	\begin{split}
		\langle T^u_{~u}(x_1) T^u_{~u}(x_2) \rangle &=
        \dfrac{1}{2}\,c_{\text{eff}}  \, \delta(u_1-u_2)^2 \, \delta^{(d-1)}(\vec{x}_1-\vec{x}_2) \, .
	\end{split}
\end{equation}
This divergence, effectively a quantum mechanical one $\delta(u=0)$ from the the frequency part of the loop integral, is identical to that appearing in the two-point function of the energy of a quantum mechanical free particle on the line. For our purposes, the most important point of \eqref{eq:tuutwopoint} is that the UV sensitivity of the spatial lattice cancels out due to the scaling behavior expressed in~\eqref{eq:nscal2}. This cancellation also occurs in the correlators of all other stress tensor components.

\subsection{Ward identities, hyperscaling, and critical exponents}
\label{ssec:hyperscaling}

In this Section, we clarify the features and peculiarities of the $N \to 0$ limit defined by~\eqref{eq:nscal2}, focusing on scale invariance, anomalies, and hyperscaling violations. We begin by examining the Ward identities and then explore how the observed critical behavior deviates from standard scaling relations, leading to a modified hyperscaling law.

We start by recalling that scale invariance would hold if the scaling relation of~\eqref{eq:ceffscald} were satisfied. To verify scale invariance explicitly, we consider the correlator involving the trace of the stress tensor, given by~\eqref{eq:scalartrace}, and two insertions of the operator $\mathcal{O}$,
\begin{equation}
\begin{aligned}
\langle T^{\mu}_{\ \mu}(x) \mathcal{O}(y_1) \mathcal{O}(y_2) \rangle &= (d-2)\left[\delta(u-u_1)\delta^{(d-1)}(\vec{x}-\vec{y}_1) + \delta(u-u_2)\delta^{(d-1)}(\vec{x}-\vec{y}_2)\right]\langle \mathcal{O}(y_1)\mathcal{O}(y_2) \rangle\,.
\end{aligned}
\end{equation}
This structure aligns with the general Ward identity for scale invariance in a $d$-dimensional conformal field theory (CFT)~\cite{DiFrancesco:1997nk},
\begin{equation}
\langle T^{\mu}_{\ \mu}(x) X \rangle = \sum_i \Delta_i \delta^d(x-x_i)\langle X \rangle\,,
\end{equation}
provided we identify the scaling dimension of $\mathcal{O}$ as $\Delta_{\mathcal{O}} = d-2$. However, as previously noted, despite the fact that $n$-point functions obey the dilatation Ward identity, the correlations of $\mathcal{O}$ imply a broken scale symmetry. This breaking of scale symmetry, a loop-level effect, is a sort of anomaly and it implies hyperscaling violation.

To illustrate this explicitly, consider the two-point function of $\mathcal{O}$ from~\eqref{eq:ooini}, which exhibits a particular hyperscaling form
\begin{equation}
\label{eq:hypscal}
G_{\mathcal{O}}\left(\alpha x_{12}, \frac{c_{\text{eff}}}{\alpha^{d-1}}\right) = \frac{1}{\alpha^{2(d-2)}} G_{\mathcal{O}}(x_{12}, c_{\text{eff}})\,.
\end{equation}
This scaling differs from the standard hyperscaling relation defined by critical exponents~\cite{Cardy_1996, PhysRevLett.56.416}.
To clarify this point we briefly review the standard definitions of critical exponents. Near a critical point, correlation functions decay exponentially with correlation length $\xi$, defined by the reduced temperature $t=(T-T_c)/T_c$, as:
\begin{equation}\label{eq:gcritscal}
G(x-x') = e^{-\frac{|x-x'|}{\xi}},\quad \xi \sim |t|^{-\nu}\,.
\end{equation}
The critical exponents $\alpha, \beta, \gamma, \delta, \nu, \eta$ characterize the behavior of physical quantities like the specific heat $C_H$, magnetization $M$, and susceptibility $\chi$ \cite{Cardy_1996},
\begin{equation}\label{eq:critexps}
\begin{aligned}
C_H &\sim |t|^{-\alpha}\,,\quad M|_{H=0}\sim |t|^{\beta}\,,\quad M|_{t=0}\sim H^{1/\delta}\,,\quad \chi \sim |t|^{-\gamma}\,,\quad G(x-x')|_{t=0}\sim |x-x'|^{-(d-2+\eta)}\,.
\end{aligned}
\end{equation}
For ordinary critical points, these exponents satisfy scaling laws, such as the hyperscaling relation
\begin{equation}
2-\alpha = d\nu\,,\quad \gamma=\nu(2-\eta)\,.
\end{equation}

Our critical point exhibits scaling in Euclidean time $\tau = i u$ rather than space, with a ``correlation time'' $\xi$ generated by a mass deformation of the description. With a mass the two-point function scales as
\begin{equation}
G_{\mathcal{O}}(x_1 - x_2) \sim \tau_{12}^2 \, e^{-m^2 \tau_{12}^2}\,\delta^{(d-1)}(\vec{x}_{12})\quad \implies \quad \xi \sim 1/m\,,\quad \nu=1\,.
\end{equation}
The critical point is at $m=0$ and the role of the reduced temperature $t$ of \eqref{eq:critexps} is played by $m$. The remaining exponents can be computed using standard methods \cite{Cardy_1996} and they are summarized in Table \ref{tab:critexp}. We note that, unlike standard Landau-Ginzburg theory (for which $\beta = \frac{1}{2}$), the one-point function of $\mathcal{O}$ shows unconventional scaling due to the $N\to 0$ limit resulting in $\beta = -1$.

\begin{table}
\centering
\begin{tblr}{
  colspec={Q[1cm]Q[1cm]Q[1cm]Q[1cm]},
  hlines,vlines,
  cells={valign=m,halign=c},
  rows={ht=1\baselineskip},
  row{1}={ht=1\baselineskip},
}
    $\alpha$ & $\beta$ & $\gamma$ & $\eta$ \\ 
    $\frac{3}{2}$ & $-1$ & $3$ & $-d$\\ 
\end{tblr}
\caption{\label{tab:critexp} Critical exponents of the $N\to 0$ limit of the free electric Carrollian vector model.}
\end{table}

Clearly, the exponents of Table \ref{tab:critexp} violate the standard hyperscaling law, modifying it to\footnote{A nonzero $\theta$ typically illustrates the presence of a ``dangerously irrelevant operator'' which implies the presence of another length scale that doesn't decouple in the IR \cite{Cardy_1996}. In our model, that extra length scale is $\ell$.}
\begin{equation}
2-\alpha = d(\nu - \theta), \quad \text{with} \quad \theta=1-\frac{1}{2d}\,.
\end{equation}
We conclude that our critical point displays anomalous scaling and hyperscaling violation, necessitating a modified scaling framework. The evolution of the critical point renormalization group flows will be explored further in Section~\ref{sec:crits}.

In the standard theory of critical phenomena, hyperscaling violation can also be diagnosed from non-trivial scaling of the partition function near criticality. Here, consider the partition function of the mass-deformed theory at inverse temperature $\beta$,
\begin{equation}
    \ln Z = \frac{NV}{a^{d-1}}\ln Z_L\,, \qquad Z_L = \frac{1}{1-e^{-\beta |m|}} \,,
\end{equation}
where we have used the fact that the theory is ultralocal, $Z_L$ refers to the partition sum of a single site, and $V$ is the spatial volume. Under the $N\to 0$ limit this becomes
\begin{equation}
    \ln Z = -\frac{V}{\ell^{d-1}}\ln (1-e^{-\beta |m|})\,,
\end{equation}
which tends to zero at zero temperature. However at finite temperature, and equipping $\beta$ with a scaling $\beta \to \alpha \beta$, we see that $\ln Z$ transforms with weight $d-1$ under dilatations, and is invariant under the combination of scale transformations along with changing the theory, $\ell \to \alpha \ell$.

Finally, we highlight a subtlety in applying conventional dimensional analysis of coupling constants in the $N \to 0$ limit defined by~\eqref{eq:nscal2}. The operator $\mathcal{O}^2$ has dimension $2(d-2)$ and so by the usual intuition ought to be marginal in $d=4$. If that operator is truly marginal, then the two-point function of the theory deformed by $\mathcal{O}^2$ would take the form
\begin{equation}
\label{eq:marginaldef}
\langle \mathcal{O}(x_1)\mathcal{O}(x_2) \rangle = 2 \, c_{\text{eff}}(1+\lambda(1-\delta \ln |u_{12}|) + O(\lambda^2)) \,u_{12}^2\, \delta^{(d-1)}(\vec{x}_{12})\,,
\end{equation}
where $\lambda$ parameterizes the marginal coupling and $\delta$ parameterizes the anomalous dimension of $\mathcal{O}$. However, as we find in Section~\ref{ssec:wilsonfisher} corrections of this form do not appear in the deformed theory. We find that under the scaling~\eqref{eq:nscal2} the free-field fixed point admits no marginal deformations. 

\section{Variations on the theme}\label{sec:crits}

\subsection{Coupling to a magnetic theory}\label{ssec:magscal}

As an interesting case study we couple the free $O(N)$ theory~\eqref{eq:freeonact} to a magnetic scalar $\chi$ \cite{Duval:2014uoa,Bagchi:2022eav,Ciambelli:2023xqk,Chen:2024voz,Ekiz:2025hdn}. The Lagrangian is given by
\begin{equation}\label{eq:lagmag}
	\mathcal{L} = \dfrac{1}{2} \!\left( \partial_u \vec{\phi} \right)^2 - \dfrac{m^2}{2} \vec{\phi}^{\,2} + \chi \partial_u Q + \dfrac{1}{2}(\vec{\nabla} \chi )^2 + \dfrac{M^2}{2} \chi^2 + \zeta  \chi \vec{\phi}^{\,2} \, ,
\end{equation}
where $\zeta$ is the scalar-Yukawa coupling. This Lagrangian is almost identical to the one considered in Section 5.1 of \cite{Cotler:2024xhb} except for the fact that the scalar here is an $O(N)$ theory. We compute simple correlation functions in this theory borrowing from the results of~\cite{Cotler:2024xhb}, implementing the $N\to 0$ limit at the end. To regulate IR divergences associated with the time direction, we study the theory at a finite inverse temperature $\beta$. Then the tree level $\chi$ and $\phi$ propagators are
\begin{equation}\label{eq:chiphitree}
	\begin{split}
		\langle \chi(x_1) \chi(x_2) \rangle_{0} &= \dfrac{1}{\beta} \,D(\vec{x}_1-\vec{x}_2) \, , \\
		\langle \phi^I(x_1) \phi_J(x_2) \rangle_{0} &= \delta^I_{~\,J} K_{\beta}(\tau_1 - \tau_2) \,\delta^{(d-1)}(\vec{x}_1-\vec{x}_2) \, ,
	\end{split} 
\end{equation}
where we have used
\begin{align}
\begin{split}
D(\vec{x}_1-\vec{x}_2) &=  \int \dfrac{d^{d-1}k}{(2\pi)^{d-1}} \dfrac{e^{i \vec{k} \cdot (\vec{x}_1-\vec{x}_2)}}{\vec{k}^2 + M^2} \, , \\
K_{\beta}(\tau_1 - \tau_2) &=
\dfrac{1}{2m} \dfrac{\cosh \left( (\frac{\beta}{2} -|\tau_1-\tau_2|)m \right) }{\sinh \left( \frac{\beta m}{2}\right) } \, .
\end{split}
\end{align}
Since the scaling~\eqref{eq:nscal2} will not affect correlators of $\partial_{\tau} Q$, we will concern ourselves with correlators of $\chi$ and the singlet operator $\mathcal{O}$.

We find the following one- and two-point functions in the $N\to 0$ limit:
\begin{align}
\langle \chi \rangle &= \dfrac{\zeta \, c_{\text{eff}}}{m} + O(\zeta^3) \, , \\
\langle \chi(\tau_1,\vec{x}_1) \chi(\tau_2,\vec{x}_2)\rangle &= \dfrac{1}{\beta}\,D(\vec{x}_1-\vec{x}_2) - \dfrac{\zeta^2}{2 \beta} c_{\text{eff}} \left( \dfrac{\partial}{\partial m^2} K_{\beta}(0) \right) \left( \dfrac{\partial}{\partial M^2} D(\vec{x}_1-\vec{x}_2) \right) + O(\zeta^4) \, , \\
\langle \mathcal{O} \rangle &= c_{\text{eff}} \, K_{\beta}(0) + 8 \zeta^2 c_{\text{eff}}\, D(\vec{0}) \int_{0}^{\beta} d\tau \, d\tau' \, K_{\beta}(\tau) \, K_{\beta}(\tau') \, K_{\beta}(\tau-\tau') \\
\nonumber & \qquad \qquad \quad \,\, + 4 \,\zeta^2\,c^2_{\text{eff}} \, \beta \, K_{\beta}(0) \, \int_{0}^{\beta} d\tau \, d^{d-1}y' \left( K_{\beta}(\tau) \right)^2 D(\vec{y}\,')\, + O(\zeta^3) \, , \\
\langle \mathcal{O}(x_1) \mathcal{O}(x_2) \rangle &=  c_{\text{eff}} \left(K_{\beta}(\tau_1-\tau_2) \right)^2 \, \delta^{(d-1)}(\vec{x}_1 - \vec{x}_2) \\
& \quad + 16 \,\zeta^2 c_{\text{eff}} \, D(\vec{0}) \, \delta^{(d-1)}(\vec{x}_{1} - \vec{x}_2) \left(\int_{0}^{\beta} d\tau \, K_{\beta}(\tau_1-\tau)  K_{\beta}(\tau-\tau_2)\right)^{\! 2}  \nonumber \\
& \quad + 8\,\zeta^2  c^2_{\text{eff}} \,  D(\vec{x}_1-\vec{x}_2) \, \int_{0}^{\beta} d\tau \, d\tau' \left( K_{\beta}(\tau_1-\tau) \right)^2 \left( K_{\beta}(\tau'-\tau_2) \right)^2\,+ O(\zeta^3)\,. \nonumber
\end{align}
The $D(\vec{0})$ divergence can be removed with a local counterterm.  Otherwise, we see the remarkable result that the $N \to 0$ limit alleviates the UV sensitivity found in~\cite{Cotler:2024xhb}, giving rise to a genuinely \textit{interacting}, renormalizable Carrollian field theory. By contrast, in~\cite{Cotler:2024xhb}, the UV sensitivity had to be managed by rescaling the scalar-Yukawa coupling $\zeta$ by the $a^{\frac{d-1}{2}}$, resulting in a theory of generalized free fields. 

\subsection{Quartic interactions}\label{ssec:wilsonfisher}

In Section~\ref{sec:nto0scalar} we analyzed the $N\to 0$ limit of the free Carrollian $O(N)$ model. It is interesting to ask what happens if we include non-trivial self interactions, such as a quartic term
\begin{equation}\label{eq:inton}
	V(\vec{\phi}\,) = \lambda (\vec{\phi}^{\,2})^2\,.
\end{equation}
Let us forgo including a mass term so that the theory is classically scale-invariant at integer $N$ in $d=4$.

We begin by computing the leading corrections to the $\langle \mathcal{O}\mathcal{O}\rangle$ two-point function in the $N\to 0$ limit at small $\lambda$. We have
\begin{align}\label{fig:phi4}
\raisebox{-0.5\height}{\includegraphics[width=6cm,height=1.5cm]{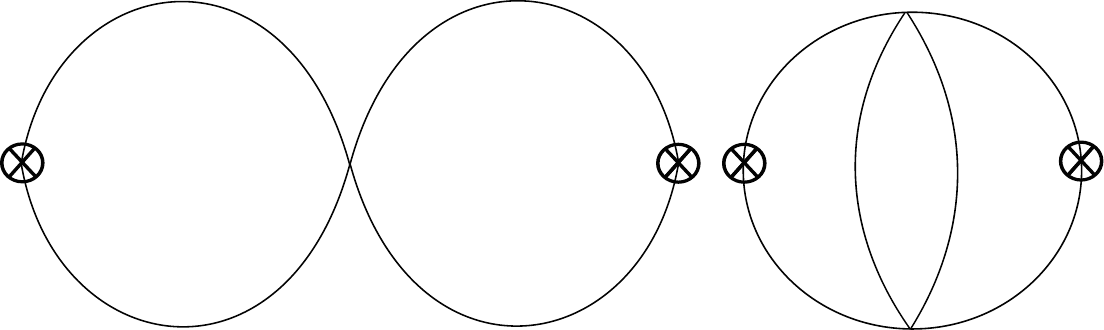}} 
\end{align}
giving
\begin{align}
\label{E:OOlambda1}
\langle \mathcal{O}(x_1) \mathcal{O}(x_2)\rangle &= 2\,c_{\text{eff}}\, u_{12}^2 \,\delta^{(d-1)}
(\vec{x}_1- \vec{x}_2) \\
& \qquad \qquad \quad + (4 \lambda \, c^2_{\text{eff}} + 8 \lambda \, c_{\text{eff}} \, \delta^{(d-1)}(0) )\, \delta^{(d-1)}(\vec{x}_1 - \vec{x}_2) I_1 (u_1,u_2) \,+O(\lambda^2)\,,\nonumber
\end{align}
where we have defined
\begin{equation}\label{eq:divint1}
	I_1 (u_1,u_2) := \int du_3 \, u^2_{13} u^2_{32} \, .
\end{equation}
The IR divergent integral $I_1$ can be regulated by working at finite temperature. The more serious divergence comes from the $\delta^{(d-1)}(0)$ term in~\eqref{E:OOlambda1}.  To make sense of it recall that in the ordinary $O(N)$ model we scale the quartic interaction as $V(\vec{\phi}\,) = \frac{\lambda}{N} (\vec{\phi}^{\,2})^2$ to get a sensible large $N$ limit.  Instead, here we will require
\begin{align}
\label{E:Vnew1}
    V(\vec{\phi}\,) = N\lambda (\vec{\phi}^{\,2})^2
\end{align}
with $\lambda$ held fixed as we take $N\to 0$.
Indeed, upon doing so, we ameliorate the divergence in~\eqref{E:OOlambda1}, and continue to find finite correlators at higher orders in perturbation theory, as well as finite correlation functions of other composite operators. For instance using~\eqref{E:Vnew1} to second order in perturbation theory we find
\begin{align}
\label{E:OOlambda2}
    \langle \mathcal{O}(x_1) \mathcal{O}(x_2)\rangle &= 2\,c_{\text{eff}} \,u_{12}^2 \,\delta^{(d-1)}
    (\vec{x}_1- \vec{x}_2) + 8 \lambda \, c_{\text{eff}}^2 \, \delta^{(d-1)}(\vec{x}_1 - \vec{x}_2) I_1 (u_1,u_2) 
    \\
    &\qquad \qquad \qquad \qquad \qquad \qquad \qquad \,\, + 32\,\lambda^2\,c_{\text{eff}}^3\,\delta^{(d-1)}(\vec{x}_1 - \vec{x}_2)\,I_2(u_1,u_2)\,+O(\lambda^3)\,,\nonumber
\end{align}
where we have defined
\begin{equation}\label{eq:divint2}
	I_2(u_1,u_2) := \int du_3 \, du_4 \, u_{13} \, u_{14} \, u^2_{34} \, u_{23} \, u_{24} \, .
\end{equation}
This integral can also be regulated at finite temperature.

There is a simple argument for the finiteness of any diagram in this scaling limit. On the one hand, every vertex contributes a factor of $N \lambda$ and a momentum-conserving delta function $\delta^{(d-1)}(0)$; hence inside each diagram we are effectively replacing $N \lambda \,\delta^{(d-1)}(0)$ by $c_{\text{eff}} \lambda$.  On the other hand, for a correlation function of singlet operators, each loop contributes a factor of $N$ and a factor $\delta^{(d-1)}(0)$ through the propagators; then we are effectively replacing $N \delta^{(d-1)}(0)$ by $ c_{\text{eff}}$. Taken all together, we find that the diagrams are UV-finite, with the exception of more ordinary renormalizable divergences.  As such, we have constructed another interacting, renormalizable Carrollian theory with finite effective central charge.

As we remarked before the scaling of~\eqref{eq:nscal2} implies an unconventional RG flow. With a mass deformation the $\langle \mathcal{O} \mathcal{O} \rangle$ correlator in the interacting theory takes the schematic form
\begin{equation}\label{eq:oolambdafull}
	\langle \mathcal{O}(x) \mathcal{O}(0) \rangle = c_{\text{eff}} \, \delta^{(d-1)}(\vec{x}) \left( 2 u^2 + \lambda \,f^{(1)}(m,u) + O(\lambda^2) \right)\,,
\end{equation}
where we have explicitly indicated the mass dependence $m$ which regulates the IR divergence of \eqref{eq:divint1} and the superscript in $f^{(1)}$ indicates the order in perturbation theory. One can explicitly work out $f^{(1)}(m,u)$ by regulating \eqref{eq:divint1} using the IR finite \eqref{eq:chiphitree} propagator,
\begin{equation}\label{eq:regdivint1}
    \int d\tau' \, (K_{\beta}(\tau_1-\tau'))^2 \, (K_{\beta}(\tau'))^2 = \dfrac{a_1 \tau_1}{m^4} + \dfrac{a_2 \tau^3_1}{m^2}+ a_3 \tau_1^5 + a_4 \tau_1^2 \, \beta^3 + a_5 \tau_1 \, \beta^4 + a_6 \, \beta^5 \, ,
\end{equation}
where the $a_i$ are some constants. \eqref{eq:regdivint1} is explicitly IR divergent in the $m \to 0, \, \beta \to \infty$ limit. The form of the monomials in \eqref{eq:regdivint1} is consistent with na\"{i}ve dimensional analysis. Under $x'=\alpha \, x$, the coupling constants change as
\begin{equation}\label{eq:rgscalecoupling}
    m' \to \dfrac{m}{\alpha} \, , ~~~~ N'\lambda' = N\lambda \, \alpha^{d-4} \quad \implies \quad \lambda' = \dfrac{\lambda}{\alpha^3} \, ,
\end{equation}
which is the scaling of a coupling of a quartic term in ordinary quantum mechanics.
\eqref{eq:regdivint1} is consistent with \eqref{eq:rgscalecoupling} as $\langle \mathcal{O} \mathcal{O} \rangle \sim c_{\text{eff}} \, u^2 \, \delta^{(d-1)}(\vec{x})$. So the quartic coupling $\lambda$ is relevant for any dimension from the Wilsonian point of view owing to \eqref{eq:rgscalecoupling}, which is rather different from traditional scalar field theory, where the quartic coupling is marginal(ly irrelevant) for $d=4$.

The $a_3$ term of \eqref{eq:regdivint1} is consistent with dimensional analysis because $ \lambda \, \tau^5_1 \to \alpha^2 \lambda \,\tau^5_1$ under $x \to \alpha \, x$. Thus, if we use~\eqref{eq:marginaldef} as our working definition of marginality on~\eqref{eq:oolambdafull}, then we expect marginality if and only if $f^{(n)}(m,u) = u^2$ for all $n$. From \eqref{eq:regdivint1}, it is clear that this cannot be true. This would imply that in general, one cannot add a coupling to the microscopic theory~\eqref{eq:freeonact} that results in a marginal deformation of the composite operator~\eqref{eq:odef}. 

Let us pause to mention an important difference between \eqref{eq:regdivint1} and conventional quantum field theories. The perturbative corrections of Fig. \ref{fig:phi4} are UV finite courtesy of \eqref{E:Vnew1}. In usual local QFTs, such loop corrections are UV divergent and they can be taken care of by adding local counterterms, which in the traditional Wilsonian paradigm corresponds to the running of coupling constants under an effective UV scale.  However, in our case, the RG running is purely due to the leading order na\"{i}ve dimensional analysis \eqref{eq:rgscalecoupling}. It is now interesting to speculate if the microscopic theory~\eqref{eq:inton} flows to a Wilson-Fisher like fixed point in the $m \to 0, \beta \to \infty$ limit for some $d$ \cite{Banerjee:2023jpi,Bagchi:2024unl}. From \eqref{eq:rgscalecoupling}, it is clear that there is no such non-trivial fixed point at weak coupling. There is no log divergent UV behaviour and thus, the conventional beta function in our case gets contributions only from the leading order dimensional analysis through \eqref{eq:rgscalecoupling}. This is consistent with the scenario in quantum mechanics where every perturbative interaction in the potential for the Schr\"{o}dinger equation is a relevant perturbation. In fact, the scaling \eqref{eq:rgscalecoupling} is identical to the scalings of a quartic quantum mechanical harmonic oscillator. Thus, despite the $N\to 0$ limit, the scaling \eqref{E:Vnew1} is effectively that of an on-site quantum mechanics.

\subsection{Coupling to Yang-Mills}\label{ssec:yangmills}

Finally we turn to non-abelian gauge theories. Specifically, we consider a massless complex scalar $\Phi^a$ coupled to a $SU(N_c+1)$ gauge field. The relativistic parent is
\begin{equation}\label{eq:ymaction}
	S_{\text{YM}} = \int du \, d^{d-1}\vec{x} \, \left( (D_{\mu} \Phi)^{\dagger} (D^{\mu}\Phi) - \dfrac{1}{4} F^a_{\mu\nu} F^{a,\mu\nu}  \right) \, , 
\end{equation}
where
\begin{equation}
	D_{\mu} \Phi^a = \partial_{\mu} \Phi^a - i g  f^{abc} A^b_{\mu} \Phi^c \, , ~~~~~ F^a_{\mu\nu} = \partial_{\mu} A^a_{\nu} - \partial_{\nu} A^a_{\mu} - i g f^{abc} A^b_{\mu} A^c_{\nu} \, .
\end{equation}
Here $g$ denotes the Yang-Mills coupling and $f^{abc}$ are the structure constants of $SU(N_c+1)$. For reasons that will be soon be clear we have taken the scalar field transforms in the adjoint representation of the gauge group and taken the rank of the gauge group to be $N_c+1$. In an $R_{\xi}$ version of Coulomb gauge the Carrollian limit of the gauge-fixed version of~\eqref{eq:ymaction} is given by \cite{Henneaux:2021yzg,Islam:2023rnc}
\begin{align}\label{eq:carymaction}
		S_{\text{YM}} &= \int \! du \, d^{d-1}\vec{x}  \Big( |\partial_u \Phi^a - i g f^{abc} A^b_u \Phi^c|^2 + \dfrac{1}{2}(\partial_u A^a_i - \partial_i A^a_u )^2  - ig f^{abc} (\partial_u A^a_i - \partial_i A^a_u) A^b_u A^c_i   \\
		& \qquad \qquad \qquad \qquad \qquad \qquad \qquad \qquad \left. - g^2 f^{abc} f^{ade} A^b_u A^c_i A^d_u A^e_i + \dfrac{1}{2 \xi} (\nabla^i A^a_i )^2 + (\text{ghosts})  \right)\,.\nonumber
\end{align}
 As usual \cite{Cotler:2024xhb}, the theory~\eqref{eq:carymaction} can be regulated by putting it on a hypercubic lattice and compactifying the imaginary time direction to a thermal circle $\tau = i u\sim \tau + \beta$. 
The tree-level propagators for the theory~\eqref{eq:carymaction} are entirely analogous to~\eqref{eq:chiphitree}, and are given by
\begin{align}\label{eq:phiatautree}
\begin{split}
    \langle \Phi^{a \dagger}(x_1) \Phi^b(x_2) \rangle_0 &= \delta^{ab} K_{\beta}(\tau_1-\tau_2) \, \delta^{(d-1)}(\vec{x}_1-\vec{x}_2)\,,
    \\
    \langle A^a_{\tau}(x_1) A^b_{\tau}(x_2) \rangle_0 &= \dfrac{\Gamma(\frac{d-3}{2})}{4 \pi^{\frac{d-1}{2}}} \dfrac{\delta^{ab}}{|\vec{x}_1-\vec{x}_2|^{d-3}} \,.
\end{split}
\end{align}
The analogue of the composite operator~\eqref{eq:odef} is
\begin{equation}\label{eq:oymdef}
	\mathcal{O}_{\text{YM}}(\tau,\vec{x}) = \,:\Phi^{a\dagger}(\tau,\vec{x}) \, \Phi_{a}(\tau,\vec{x}): 
\end{equation}
which is a singlet under $SU(N_c+1)$. At vanishing coupling the two-point function of~\eqref{eq:oymdef} is
\begin{equation}\label{eq:ooymini}
		\langle \mathcal{O}_{\text{YM}}(\tau_1,\vec{x}_1) \mathcal{O}_{\text{YM}}(\tau_2,\vec{x}_2) \rangle_0 		= \left( N^2_c + 2 N_c \right) \left( K_{\beta}(\tau_1-\tau_2)\right)^2 \delta^{(d-1)}(\vec{x}_1-\vec{x}_2) \, \delta^{(d-1)}(\vec{0}) \, ,
\end{equation}
where $N_c^2 + 2 N_c = (N_c + 1)^2 - 1$ is the number of independent matrix degrees of freedom running around the loop. Since we considered $SU(N_c+1)$ instead of $SU(N_c)$, the linear term in $N_c$ dominates over the quadratic term as $N_c\to 0$. Thus, we can use the same scaling we used for the vector model, analytically continuing $N_c$ to non-integer values $N_c = \left( \frac{a}{\ell}\right)^{d-1}$ and holding the length scale $\ell$ fixed while sending $a\to 0$. This results in a model with an effective central charge
\begin{equation}\label{eq:nscalym}
	\boxed{c_{\text{YM,eff}} = \frac{1}{\ell^{d-1}} }
\end{equation}
Implementing this scaling in~\eqref{eq:ooymini} we find 
\begin{equation}\label{eq:ooym}
	\langle \mathcal{O}_{\text{YM}}(\tau_1,\vec{x}_1) \mathcal{O}_{\text{YM}}(\tau_2,\vec{x}_2) \rangle_0 = 2 \,c_{\text{YM,eff}} \left( K_{\beta}(\tau_1-\tau_2)\right)^2 \, \delta^{(d-1)}(\vec{x}_1-\vec{x}_2) \, .
\end{equation}
With our chosen scaling, the matrix degrees of freedom effectively reduce to vector degrees of freedom at small $N_c$, allowing us to borrow both calculations and intuitions from our analysis of the vector model. Note that this $N_c\to 0$ limit would not result in a finite theory if $\Phi$ transformed in the vector representation.

The first perturbative corrections to~\eqref{eq:ooym} arise from the cubic coupling to $A_{\tau}$ and are given by the diagrams
\begin{align}
\raisebox{-0.5\height}{\includegraphics[width=6cm,height=1.5cm]{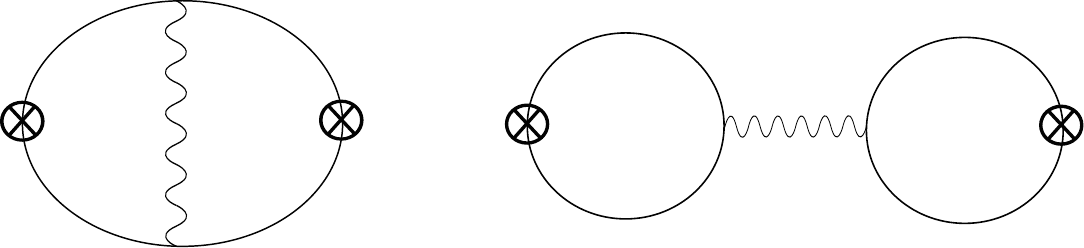}} 
\end{align}
The coupling is identical to the scalar-Yukawa coupling of the magnetic theory considered in Section~\ref{ssec:magscal} except that the role of $\chi$ is played by $A_{\tau}$. Essentially identical diagrams arise in the scalar-Yukawa coupling of the magnetic theory; the only salient difference in the Yang-Mills setting are factors associated with the rank of the gauge group that we work out below.  As such, we will just track the lattice dependence of the diagram. The diagrams each have four scalar propagators and two vertices, and the first gives
\begin{equation}\label{eq:ooymg21}
		\langle \mathcal{O}_{\text{YM}}(\tau_1,\vec{x}_1) \mathcal{O}_{\text{YM}}(\tau_2,\vec{x}_2) \rangle_{O(g^2)} 
        \sim 2  \,c_{\text{YM,eff}} \, g^2 \delta^{(d-1)}(\vec{x}_1 - \vec{x}_2)\,.
\end{equation}
The second diagram is identically zero in the adjoint representation because the contractions of the complex scalar field result in a proportionality factor of $f^{daa}$ which vanishes because $f^{abc}$ is totally antisymmetric.  We observe that the lattice-sensitive UV scale drops out without us needing to explicitly rescale the Yang-Mills gauge coupling $g$.

We can similarly consider the one-point function of $\mathcal{O}_{\text{YM}}$, which at $O(g^2)$ is
\begin{align}
\raisebox{-0.5\height}{\includegraphics[width=6cm,height=1.5cm]{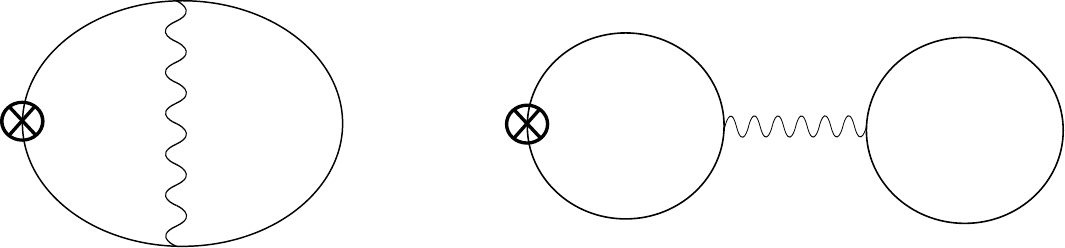}} 
\end{align}
The contributions of these diagrams mirror those in the scalar-Yukawa setting, except for the rank factors from the gauge group. The diagrams have three scalar propagators and two vertices, and the first diagram goes as
\begin{equation}
		\langle \mathcal{O}_{\text{YM}} \rangle_{O(g^2)} 
		 \sim 2 \, c_{\text{YM,eff}} \, g^2 \, .
\end{equation}
The second diagram vanishes due to a factor of $f^{daa}$ which is zero by antisymmetry.  As before, we have ameliorated the UV sensitivity without rescaling $g$.

As a last example, we consider $\langle A_\tau A_\tau \rangle$.  Its perturbative corrections are given by
\begin{align}
\raisebox{-0.5\height}{\includegraphics[width=6cm,height=1.5cm]{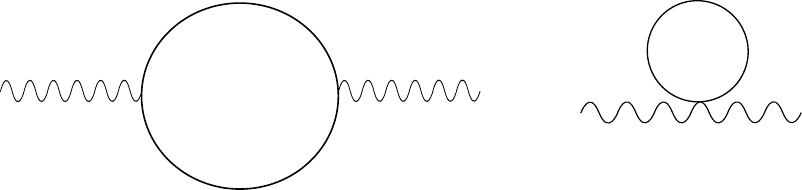}} \,,
\end{align}
along with the same diagrams where the adjoint scalar is replaced by the ghosts. The results of these diagrams are identical to those computed in~\cite{Cotler:2024xhb}, and so here we will be content with working out their UV sensitivity. The first diagram has two scalar propagators and two vertices, giving us
\begin{equation}
    \sim f^{cae} f^{cbe} g^2 \, \left( \int d^{d-1}x' \right)^2 \, \left( \delta^{(d-1)}(\vec{0})\right)^2 \sim \delta^{ab} (N^2_c + 2 N_c) g^2 \to 0 
\end{equation}
The second diagram has one scalar propagator and one vertex, giving
\begin{equation}
    \sim f^{cae} f^{cbe} g^2 \, \left( \int d^{d-1}x' \right) \, \left( \delta^{(d-1)}(\vec{0})\right) \sim \delta^{ab} (N^2_c + 2 N_c) g^2 \to 0 
\end{equation}
which also scales to zero.  As such, the would-be UV sensitivities present in these diagrams are alleviated by our $N\to 0$ limit. 

\subsection*{Acknowledgements}

It is a pleasure to thank A.~Bagchi, L.~Freidel, and R.~Ruzziconi for enlightening discussions. JC is supported by the Simons Collaboration on Celestial Holography. PD and KJ are supported in part by an NSERC Discovery grant.

\bibliography{refs}
\bibliographystyle{JHEP}

\end{document}